\documentclass[12pt]{article}
\usepackage{amsfonts}
\usepackage{amsfonts}
\usepackage{amsmath,amssymb}
\usepackage{mathrsfs}
\usepackage{bbm}

\makeatletter\renewcommand{\section}{\@startsection
{section}{1}{\z@}{-3.5ex plus -1ex minus
    -.2ex}{2.3ex plus .2ex}{\bf }}

\makeatletter \@addtoreset{equation}{section}

\usepackage[vcentermath,enableskew]{youngtab}

\newcommand{\unit}{\mathbbm{1}}   
\newcommand{\eff}{{\mathrm{eff}}}   			
\newcommand{\ah}{\hat{a}}
\newcommand{\sU}{\mathsf{U}}     			

\newcommand{\sSU}{\mathsf{SU}}

\newcommand{\sEnd}{\mathsf{End}\,}
\newcommand{\CCL}{\mathscr{L}}    
\newcommand{\CC}{\mathcal{C}}    
\newcommand{\CCD}{\mathscr{D}}    
\newcommand{\CF}{\mathcal{F}}    
\newcommand{\CH}{\mathcal{H}}    
\newcommand{\CO}{\mathcal{O}}    

\newcommand{\FR}{\mathbbm{R}}     
\newcommand{\FC}{\mathbbm{C}}     
\newcommand{\CPP}{{\mathbbm{C}P}}    
\newcommand{\RZ}{\mathbbm{Z}}     
\newcommand{\dd}{\mathrm{d}}     
\newcommand{\dpar}{\partial}     
\newcommand{\diag}{{\mathrm{diag}}}     
\newcommand{\de}{\mathrm{e}}     
\newcommand{\di}{\mathrm{i}}     
\newcommand{\bz}{{\bar{z}}}     
\newcommand{\eps}{{\varepsilon}}     
\newcommand{\eand}{{~~~\mbox{and}~~~}}     
\newcommand{\ewith}{{~~~\mbox{with}~~~}}     
\newcommand{\der}[1]{\frac{\dpar}{\dpar #1}}   
\newcommand{\tr}{\,\mathrm{tr}\,}     

\newcommand{\remark}[1]{}     
\newcommand{\vac}{|0\rangle}
\newcommand{\cav}{\langle 0|}

\begin{document}
\begin{flushright}
  DIAS-STP-07-16
\end{flushright}
\begin{center}
\Large A Multitrace Matrix Model from Fuzzy Scalar Field Theory\footnote{Talk given by CS at the
International Workshop ``Supersymmetries and Quantum Symmetries'' (SQS'07),
Bogoliubov Laboratory of Theoretical Physics, JINR, Dubna, July 30 -- August 4 2007.} \\[0.2cm]
\large Denjoe O'Connor and Christian S\"{a}mann\\[0.2cm]
\normalsize
{\em School of Theoretical Physics\\
Dublin Institute for Advanced Studies\\
10 Burlington Road, Dublin 4, Ireland}\\[2mm]
{\ttfamily denjoe, csamann@stp.dias.ie}
\end{center}

\begin{quote}
We present the analytical approach to scalar field theory on the fuzzy sphere which has been developed in hep-th/0706.2493. This approach is based on considering a perturbative expansion of the kinetic term in the partition function. After truncating this expansion at second order, one arrives at a multitrace matrix model, which allows for an application of the saddle-point method. The results are in agreement with the numerical findings in the literature.
\end{quote}

\section{Introduction}

It seems quite natural to expect that as one approaches the Planck scale, one has to replace the smooth structure of spacetime by some form of quantized geometry. The usual quantization procedure as well as string theory suggest that the first modifications that should be encountered are noncommutative geometries with constant deformation parameters. Besides the well-known Moyal-plane $\FR^2_\theta$ and its $2d$-dimensional generalizations with their quantized function algebras based on $[\hat{x}^\mu,\hat{x}^\nu]=\di\theta^{\mu\nu}$, the so-called {\em fuzzy geometries} have received more and more attention recently. 

A fuzzy geometry is essentially a noncommutative deformation of a Riemannian manifold which come with a Laplace operator with a discrete spectrum. Essentially, one truncates this spectrum and deforms the algebra of the corresponding truncated set of eigenfunctions to achieve closure under multiplication. The most prominent example of such a space is the fuzzy sphere \cite{Berezin:1974du} with a quantized function algebra based on the relation $[\hat{x}^i,\hat{x}^j]\sim\theta\di \eps_{ijk}\hat{x}^k$. 

Since fuzzy spaces are described by function algebras with finitely many degrees of freedom, they might prove useful as regulators for quantum field theories. After calculating the path integral depending explicitly on the deformation parameter $\theta$, one should be able to recover the commutative path integral  in a certain limit $\theta\rightarrow 0$. This regularization procedure would have several advantages over the lattice approach as, for example, it preserves a number of continuous symmetries and is not expected to suffer from the fermion doubling problem. Furthermore, numerical simulations are easily performed within this framework. 

A possible obstacle to using fuzzy geometries as regulators has been pointed out by various authors: the na{\"i}ve application of this procedure yields the wrong commutative limits in the case of fuzzy scalar field theory \cite{Vaidya:2001bt}, as is also suggested by numerical studies of the phase diagram \cite{Martin:2004un}. However, modifications of the action of the theory have been proposed \cite{Dolan:2001gn} and for scrutinizing them, an analytical handle on the partition function of this theory is desirable. In \cite{O'Connor:2007ea}, such an approach was developed and we will review the approach and summarize the findings in the following.

\section{The fuzzy sphere and further fuzzy geometries}

To begin, let us briefly review the construction of a number of fuzzy geometries. First, recall that the complex space $\FC^n\cong \FR^{2n}$ is rendered noncommutative by replacing the complex coordinates $z_\alpha,\bz_\beta$ with the creation and annihilation operators of $n$ harmonic oscillators satisfying the algebra $[\hat{a}_\alpha,\hat{a}^\dagger_\beta]=\delta_{\alpha\beta}$. Functions become thus linear operators on the infinite-dimensional Fock space $\CF$ generated by the creation operators from the vacuum $|0\rangle$. By normalizing the coordinates $z_\alpha\rightarrow z_\alpha/|z|$, we descend from $\FC^n$ to the sphere $S^{2n-1}$. To descend further to $\CPP^{n-1}$, we can use the Hopf fibration
\begin{equation}
 1 \ \rightarrow\  \sU(1) \ \rightarrow\  S^{2n-1} \ \rightarrow\  \CPP^{n-1} \ \rightarrow\  1~,
\end{equation}
which tells us that functions on $\CPP^{n-1}$ are obtained from the functions on $S^{2n-1}$ by factoring out a $\sU(1)$-action. We can take this action to be $z_\alpha\mapsto\de^{\di\varphi} z_\alpha$, and thus we see that functions on $\CPP^{n-1}$ are built from monomials containing an equal number of $z_\alpha$ and $\bz_\alpha$. As a basis for these functions, we use 
\begin{equation}
z_{\alpha_1}\ldots z_{\alpha_L}\bz_{\beta_1}\ldots \bz_{\beta_L}~, 
\end{equation}
which under the above quantization prescription turns into the operator basis \cite{Dolan:2006tx}
\begin{equation}\label{opbasis}
\ah^\dagger_{\alpha_1}\ldots \ah_{\alpha_L}^\dagger\vac\cav \ah_{\beta_1}\ldots \ah_{\beta_L}~.
\end{equation}
This basis spans the space of linear operators acting in the $L$-particle Hilbert space $\CH_L$ of the above Fock space $\CF$. We have $\dim(\CH_L)=\frac{(n-1)\ldots (n+L)}{L!}$, and in particular, for the sphere $S^2\cong \CPP^1$, $\dim(\CH_L)=L+1$. Functions on the fuzzy sphere can thus be encoded in $(L+1)^2$-dimensional matrices; real functions correspond to hermitian matrices. Note that this map between monomials and operators also gives a direct quantization prescription for the function algebra of a projective algebraic variety as discussed in \cite{Saemann:2006gf}. 

Note also that the quantization of $\CPP^n$ as described above corresponds to the geometric quantization (Toeplitz quantization) of $\CPP^n$ with the quantum line bundle $\CCL=\CO(1)^{\otimes L}\cong \CO(L)$: Recall that this quantization procedure consists of replacing the algebra of smooth functions on a projective algebraic variety, $C^\infty(M)$, with the endomorphisms of sections $\sEnd(\Gamma(M,\CCL))$ of the quantum line bundle $\CCL$. As the set of sections $\Gamma(\CPP^1,\CO(L))$ is spanned by $z_{\alpha_1}\ldots z_{\alpha_L}$, the quantized algebra is spanned by the operators $z_{\alpha_1}\ldots z_{\alpha_L}\der{z_{\beta_1}}\ldots \der{z_{\beta_L}}$ and it is thus equivalent to the operator basis \eqref{opbasis}.

To capture the topology of the space we are quantizing, it is crucial to specify an additional structure, which is usually taken to be a Dirac operator in noncommutative geometry. For our purposes, however, it is sufficient to define a Laplace operator. Its definition is most easily gleaned from an alternative, group theoretic point of view.

Consider a finite dimensional, irreducible representation $\rho$ of $\sSU(n)$, extended to a representation of $\sU(n)$. Such a representation can be labeled by a Dynkin diagram endowed with Dynkin labels $a_1,\ldots ,a_{n-1}$:
\begin{equation}\label{Dynkin}
\begin{picture}(180,40)(0,-15)
\put(20.0,0.0){\circle{10}}
\put(25.0,0.0){\line(1,0){40}}
\put(70.0,0.0){\circle{10}}
\put(75.0,0.0){\line(1,0){40}}
\put(125.0,0.0){\makebox(0,0)[c]{$\ldots $}}
\put(135.0,0.0){\line(1,0){40}}
\put(180.0,0.0){\circle{10}}
\put(20.0,12.0){\makebox(0,0)[c]{$a_1$}}
\put(70.0,12.0){\makebox(0,0)[c]{$a_2$}}
\put(180.0,12.0){\makebox(0,0)[c]{$a_{n-1}$}}
\put(-5.0,-12.0){\makebox(0,0)[c]{$\sU(1)$}}
\put(45.0,-12.0){\makebox(0,0)[c]{$\sU(1)$}}
\put(95.0,-12.0){\makebox(0,0)[c]{$\sU(1)$}}
\put(155.0,-12.0){\makebox(0,0)[c]{$\sU(1)$}}
\put(205.0,-12.0){\makebox(0,0)[c]{$\sU(1)$}}
\end{picture}
\end{equation}
where we have arranged the Cartan generators of the maximal torus $\sU(1)^{\times n}$ around the corresponding roots $\vec{\alpha}_i$ corresponding to the Dynkin labels $a_i$. To every simple root, we have a pair of raising and lowering operators, $E^\pm_{\vec{\alpha}_i}$, and the Dynkin label indicates the highest non-trivial action of the lowering operator on the highest weight state $|\mu\rangle$ of this representation:
\begin{equation}
 (E^-_{\vec{\alpha}_i})^{a_i}|\mu\rangle\ \neq\ 0\ =\ (E^-_{\vec{\alpha}_i})^{a_i+1}|\mu\rangle~.
\end{equation}
Putting Dynkin labels $a_i$ to zero enlarges the isotropy group of the highest weight state $|\mu\rangle$ in $\rho$, as the generators $E^\pm_{\vec{\alpha}_i}$ now leave $|\mu\rangle$ invariant and combine the neighboring $\sU(n)\times\sU(m)$ in the diagram \eqref{Dynkin} to $\sU(n+m)$. One can now show that there is a one-to-one correspondence between points $p$ on the coset space $\sU(n)/(\sU(m_1)\times\ldots \times \sU(m_k))$ and coherent states $|p\rangle$ in a representation $\rho$ in which the isotropy group of the highest weight state $|\mu\rangle$ is $\sU(m_1)\times\ldots \times \sU(m_k)$. 

An obvious quantization prescription is now to associate an operator $\hat{f}$ to a function $f(p)$ according to
\begin{equation}\label{quantization}
 f(p)\ =\ \langle p|\hat{f}|p\rangle~,
\end{equation}
where $|p\rangle$ is the coherent state in $\rho$ corresponding to the point $p$. In this way, one can find quantizations for all complex flagmanifolds and their supersymmetric extensions as shown in \cite{Murray:2006pi}.

Recall also that the Young diagram corresponding to the representation $\rho(a_1,\ldots ,a_{n-1})$ is given by
\begin{equation}
n-1\Bigg\{\overbrace{\hbox{\footnotesize$\yng(8,4,2)$}}^{a_1+\ldots +a_{n-1}}
\end{equation}
which in the case of $S^2\cong \CPP^1 \cong \sU(2)/(\sU(1)\times\sU(1))$ just becomes $\rho(a_1)$ with $a_1=L$ or
\begin{equation}
\overbrace{\hbox{\footnotesize$\yng(4)$}}^{L}~~\ \cong\ ~~\mathrm{span}(~\hat{a}^\dagger_{\alpha_1}\ldots \hat{a}^\dagger_{\alpha_L}|0\rangle~)~
\end{equation}
A function is therefore mapped via \eqref{quantization} to a linear operator,
\begin{equation}
\hat{f}\in\overbrace{\hbox{\footnotesize$\yng(4)$}}^{L}\otimes\overbrace{\hbox{\footnotesize$\yng(4)$}}^{L}~\ \cong\ ~\mathrm{span}(~ \hat{a}^\dagger_{\alpha_1}\ldots \hat{a}^\dagger_{\alpha_L}|0\rangle\langle 0|\ah_{\beta_1}\ldots \ah_{\beta_L}~)~,
\end{equation}
and we arrive again at the above quantization procedure. 

The representation of the generators of the isometries on $\CPP^1\cong \sSU(2)/\sU(1)$ on states in $\rho$ is the usual Schwinger representation, $\hat{L}_i=\ah^\dagger_\alpha \sigma_{\alpha\beta}^i\ah_\beta$. The Laplace operator is then the square of this action on functions, which is also the second Casimir operator in $\rho$:
\begin{equation}
 \Delta f \ \rightarrow\  \hat{\Delta}\hat{f}\ :=\ \hat{C}_2\hat{f}\ =\ [\hat{L}_i,[\hat{L}_i,\hat{f}]]~.
\end{equation}
The requirement that the measure involved in the integration procedure is invariant under the isometries of $\CPP^1$ as well as the usual rule of partial integration ($\int_M \dd \alpha=\int_{\dpar M} \alpha=0$, as the spaces $M$ under consideration are compact) and the volume formula on the sphere force us to define integration according to $\int_{S^2} \dd A~f \ \rightarrow\  \frac{4\pi R^2}{N}\tr(\hat{f})$.

We have now a complete quantization of the space $\CPP^1$ at hand, and we can easily translate a commutative scalar field theory to the noncommutative setting. From now on, we will omit hats over operators for convenience. Also, in accordance with the standard nomenclature of matrix models, we will label the size of the matrices encoding functions by $N=L+1$.

\section{Fuzzy scalar field theory}

With the prescription of the last section, we immediately arrive at the following action for scalar $\phi^4$-theory on the fuzzy sphere:
\begin{equation}\label{ActionGen}
S=\gamma\tr\left(\frac{a}{R^2} \Phi C_2\Phi+r\,\Phi^2+g\,\Phi^4\right)\ =\ \gamma\tr\left(-\frac{a}{2R^2}[L_i,\Phi][L_i,\Phi]+r\,\Phi^2+g\,\Phi^4\right)~,
\end{equation}
where we introduced the shorthand notation $\gamma=\frac{4\pi R^2}{N}$. For simplicity, we will put $R=1$ in the following. The partition function of the model reads as 
\begin{equation}\label{PartitionGen}
Z \ =\ \int \dd \mu_D(\Phi)~ \de^{-\beta S}\ =\ \int \dd \mu_D(\Phi)~ \de^{-\beta\gamma\tr\left(-\frac{a}{2}[L_i,\Phi][L_i,\Phi]+r\,\Phi^2+g\,\Phi^4\right)}~,
\end{equation}
where the measure $\dd \mu_D(\Phi)$ denotes the Dyson measure on the set of hermitian matrices of dimension $N\times N$, $\dd \mu_D(\Phi):=\prod_{i\leq j}\dd\Re(\Phi_{ij})\prod_{i<j}\dd\Im(\Phi_{ij})$.

It is well-known that the pure matrix model limit $a=0$ of this model has a third order phase transition in the large $N$ limit at $g\ =\ \frac{\gamma}{4N} r^2\ =\ \frac{\pi }{N^2} r^2$ for $r\leq 0$. At this parabola in the left half of the $r$-$g$-plane, the double well potential becomes sufficiently deep for the support of the eigenvalue density to split into two disjoint pieces.

Numerical studies on the lattice \cite{Loinaz:1997az} of the planar commutative limit of this model, 
\begin{equation}\label{planarlimit}
 Z \ =\ \int \CCD \phi~ \de^{-\int\dd^2 x \frac{1}{2}(\nabla \phi)^2+r \phi^2+g \phi^4}~,
\end{equation}
indicate a second-order phase transition for $g/r\approx -10.24$, i.e.\ at a line in the $r$-$g$-plane.

Interestingly, numerical studies \cite{Martin:2004un} of the fuzzy scalar field theory \eqref{ActionGen} find a combination of both phase transitions with a triple point at which the parabola meets the line. This seems at least to suggest a contradiction to the idea that the fuzzy sphere can be used as a regulator for scalar quantum field theory on the plane. After gaining an analytical handle on the phase diagram, we will therefore apply our technique to study modifications proposed to cure this problem and we will see that the triple point is shifted off to infinity; this is an indication for an improved situation as the phase diagram turns into the one of \eqref{planarlimit}.

Let us briefly recall why scalar field theory on the fuzzy sphere is more difficult than the more common matrix models. First, let us consider again the matrix model limit $a=0$, which is exactly solvable \cite{Brezin:1977sv}. Using the decomposition $\Phi=\Omega\Lambda\Omega^\dagger$, where $\Lambda=\diag(\lambda_1,\ldots ,¸\lambda_N)$ and $\Omega$ is a unitary matrix, the dependence on $\Omega$ drops out from the action due to cyclicity of the trace. The partition function \eqref{PartitionGen} turns into the one of the eigenvalue model
\begin{equation}
\begin{aligned}
Z_{a=0}&\ =\ \int \CCD\lambda~ \Delta^2(\Lambda) \int \dd \mu_H(\Omega)~ \de^{-\beta\gamma\tr\left(r\sum_i \lambda_i^2+g\sum_i \lambda_i^4\right)}\\&\ =\ \int \CCD\lambda ~ \de^{-2 \sum_{i>j}\ln|\lambda_i-\lambda_j|-\beta\gamma\left(r\sum_i \lambda_i^2+g\sum_i \lambda_i^4\right)}~,
\end{aligned}
\end{equation}
where $\CCD\lambda:=\prod_{i=1}^N \dd \lambda_i$, $\dd \mu_H(\Omega)$ is the Haar measure on $\sU(N)$ and $\Delta(\Lambda)$ is the Vandermonde determinant $\Delta(\Lambda)\ :=\ \det ([\lambda_i^{j-1}]_{ij})\ =\ \prod_{i>j} (\lambda_i-\lambda_j)$. From here, one can continue with various techniques to evaluate $Z$, as, e.g., the saddle-point approximation or the method of orthogonal polynomials. The latter even yields an exact result for $Z$ at finite $N$. However, $\Omega$ does not commute with the external matrices $L_i$ appearing in our model \eqref{ActionGen} and thus this method is not directly applicable in our case.

More generally, there is a solution for hermitian matrix models of the form $S=\tr(V_1(A\Phi)+V_2(\Phi))$ with a single external matrix $A$ as shown in \cite{DiFrancesco:1992cn}: One can use a character expansion together with the orthogonality relation for elements of $\sU(N)$ in an irreducible representations $\rho$, 
\begin{equation}\label{orthogonalityrelation}
\int \dd \mu_H(\Omega) \chi_\rho(A\Omega^\dagger \Lambda \Omega)\ =\ \frac{1}{\dim(\rho)}\chi_\rho(A)\chi_\rho(\Lambda)~,
\end{equation}
to arrive at a closed formula for the partition function. Unfortunately, having three external matrices as in our action \eqref{ActionGen} renders the character expansion so complicated that it is essentially useless for our purposes.

\section{Calculating the phase diagram}

Before performing the perturbative calculation, we can employ symmetry arguments to make some statements about the expected results. First of all, we recall that the Dyson measure is invariant under the adjoint action by a unitary matrix $\Omega$: $\dd \mu_D(\Phi)\ =\ \dd \mu_D(\Omega\Phi \Omega^\dagger)$. This implies that we can replace the action \eqref{ActionGen}, which is not invariant under this action, by an effective action $S_\eff$ with this invariance under the functional integral:
\begin{equation}
\int \dd\mu_D(\Phi)~ \de^{-S}\ =\ \int \dd\mu_D(\Phi)~ \de^{-S_\eff}~,~~~
\de^{-S_\eff[\Phi]}\ =\ \frac{1}{\mathrm{vol}(\sU(N))}\int \dd \mu_H(\Omega)~ \de^{-S[\Omega \Phi \Omega^\dagger]}~.
\end{equation}
Because of this invariance, the effective action has to be of the form
\begin{equation}\label{patternmultitrace}
 S_\eff\ =\ \sum_n s_n \tr(\Phi^n)+\sum_{n,m} s_{nm} \tr(\Phi^n)\tr(\Phi^m)+\sum_{n,m,k}s_{nmk}\tr(\Phi^n)\tr(\Phi^m)\tr(\Phi^k)+\ldots~,
\end{equation}
which can now be trivially recast into an eigenvalue model $S_\eff(\Lambda)$.

Furthermore, symmetry arguments severely restrict the form in which the kinetic term can appear in the effective action $S_\eff(\Lambda)$. Since $C_2\unit=0$, the kinetic term $\tr(\Phi C_2\Phi)$ can only depend on the difference of eigenvalues $\lambda_i-\lambda_j$. Because of the $\RZ_2$-symmetry $\lambda_i\rightarrow -\lambda_i$, we have to take this difference to an even power and the permutation symmetry between the eigenvalues demands that we sum over all $i>j$. Altogether we arrive at 
\begin{equation}\label{lambdapattern}
 S_{\eff}(\Lambda)-V(\Lambda)\ =\ \sum_{k,m_1,n_1,\ldots ,m_k,n_k} \xi_{(m_1,n_1)\ldots (m_k,n_k)}\Xi_{2m_1}^{n_1}\ldots\Xi_{2m_k}^{n_k}~,
\end{equation}
where 
\begin{equation}
 \Xi_{2m}^{n}\ :=\ \big(\Xi_{2m}\big)^{n}\ :=\ \Big(\sum_{i>j}(\lambda_i-\lambda_j)^{2m}\Big)^n~.
\end{equation}

The idea of treating the kinetic term perturbatively is known as the hopping parameter or high-temperature expansion and this technique has been successfully used to analyze two-dimensional scalar $\phi^4$-theory on the lattice, see e.g.\ \cite{Luscher:1987ay} and references therein. In our case, it will allow us to treat the partition function in principle exactly for any truncation of the perturbative series.

We will restrict our considerations to an expansion of the kinetic term up to second order:
\begin{equation}\label{3.1}
\de^{\beta\gamma a\Phi^aK_{ab}\Phi^b}\ =\ 1+\beta\gamma a \Phi^aK_{ab}\Phi^b+\frac{\beta^2\gamma^2 a^2}{2} \Phi^aK_{ab}\Phi^b\,\Phi^cK_{cd}\Phi^d+\CO(a^3)~,
\end{equation}
where
\begin{equation}
K_{ab}\ =\ \tr([L_i,\tau_a][L_i,\tau_b])\eand \Phi\ =\ \Phi^\mu\tau_\mu\ =\ \Phi^0\frac{\unit_N}{N}+\Phi^a\tau_a
\end{equation}
and $\tau_a$ are the Gell-Mann matrices of $\sSU(N)$ normalized according to $\tr(\tau_a\tau_b)=\delta_{ab}$. To integrate over the angular variables, we have to compute 
\begin{equation}
 \int \dd\mu_H(\Omega)K_{ab}\tr(\tau^a\Omega \Lambda\Omega^\dagger)\tr(\tau^b\Omega \Lambda\Omega^\dagger)\ewith
\Phi^a\ =\ \tr(\tau^a\Omega \Lambda\Omega^\dagger)~.
\end{equation}
Having in mind the orthogonality relation \eqref{orthogonalityrelation}, we rewrite 
\begin{equation}
\begin{aligned}
\tr\left( (\tau^a\Omega \Lambda\Omega^\dagger)\otimes(\tau^b\Omega \Lambda\Omega^\dagger)\right)
\ =\ 
\tr\left( (\tau^a\otimes\tau^b)(\Omega\otimes\Omega)(\Lambda\otimes \Lambda)(\Omega^\dagger\otimes \Omega^\dagger)\right)~. 
\end{aligned}
\end{equation}
Splitting the tensor product ${\raisebox{1mm}{\tiny $\yng(1)$}}\otimes{\raisebox{1mm}{\tiny $\yng(1)$}}$ into irreducible representations $\rho$, we arrive at the formula
\begin{equation}
\int\dd\mu_H(\Omega)K_{ab}\Phi^a\Phi^b\ =\ K_{ab}\sum_\rho \tfrac{1}{\dim(\rho)}\tr_{\rho}(\tau^a\otimes \tau^b)\tr_{\rho}(\Lambda\otimes \Lambda)~,
\end{equation}
where $\tr_{\rho}$ denotes the trace in the irreducible representation specified by $\rho$. The explicit expressions for $\tr_\rho$ in all relevant representations are found in \cite{O'Connor:2007ea}, and using these, one straightforwardly computes that, in accordance with \eqref{lambdapattern},
\begin{equation*}
 \int \dd \mu_H(\Phi)K_{ab}\Phi^a\Phi^b\ =\  \frac{N}{2}\sum_{i>j}(\lambda_i-\lambda_j)^2\ =\ \int \dd \mu_H(\Phi)N^2\left(\tr(\Phi^2)-\frac{1}{N}\tr(\Phi)^2\right)~.
\end{equation*}

At second order, we obtain after a more tedious but also straightforward calculation a more complicated result, which is again of the form predicted in \eqref{lambdapattern}. Instead of presenting it here, let us directly jump to the full large $N$ limit of the model. For this, we re-exponentiate the contributions found perturbatively into the effective action. Neglecting terms which are subdominant in $N$, we arrive at the following eigenvalue model:
\begin{equation*}
S\ =\ \gamma\sum_i\left(r\lambda_i^2+g\lambda_i^4\right)+\gamma\sum_{i>j}\left(-\tfrac{a}{2}N(\lambda_i-\lambda_j)^2+\tfrac{\gamma a^2}{4}N^2(\lambda_i-\lambda_j)^4-\tfrac{2}{\gamma}\ln|\lambda_i-\lambda_j|\right)~.
\end{equation*}
The transition to continuous variables is performed as usual by rescaling $\lambda_i\rightarrow\lambda(\tfrac{i}{N})=\lambda(x)$ with $0<x\leq 1$ and turning the sums into integrals: $\sum_{i=1}^N\rightarrow N\int_0^1\dd x$. Moreover, a common power of $N$ has to be factored out from every term in the action. This power is determined by the logarithmic term to be $N^2$ and yields a rescaling,
\begin{equation}
a\ =\ N^{\theta_a}\tilde{a}~,~~~r\ =\ N^{\theta_r}\tilde{r}~,~~~g\ =\ N^{\theta_g}\tilde{g}\eand\lambda(x)\ =\ N^{\theta_\lambda}\tilde{\lambda}(x)~,
\end{equation}
with values for $\theta_g$ and $\theta_r$ which remarkably are found to be consistent with the\linebreak ones obtained numerically in \cite{Martin:2004un}. Altogether, we have now the partition function $Z=\int\CCD \lambda~\exp(-N^2\tilde{S})$ with 
\begin{equation}
\begin{aligned}
\tilde{S}\ =\ 4\pi\int_0^1\dd x \Big(\tilde{r}\tilde{\lambda}^2(x)+&\tilde{g}\tilde{\lambda}^4(x)+\int_0^1\dd y \Big(-\tfrac{\tilde{a}}{4}(\tilde{\lambda}(x)-\tilde{\lambda}(y))^2\\&+\tfrac{4\pi\tilde{a}^2}{8}(\tilde{\lambda}(x)-\tilde{\lambda}(y))^4-\tfrac{1}{4\pi}\ln |\tilde{\lambda}(x)-\tilde{\lambda}(y)|\Big)\Big)~,
\end{aligned}
\end{equation}
which we can evaluate using the saddle point approximation. For this, we introduce as usual the eigenvalue density $u(\tilde{\lambda})=\dd x/\dd \tilde{\lambda}$ and follow the canonical procedure for determining $u(\tilde{\lambda})$ (see e.g.\ \cite{Brezin:1977sv}, cf.\ \cite{Das:1989fq}), which yields in the single cut regime
\begin{equation}
 u(\tilde{\lambda})\ =\ \left(4\tilde{r}-\tilde{a}+12\pi\tilde{a}^2c_2+4\left(\tilde{g}+\tfrac{\pi\tilde{a}^2}{2}\right)\delta^2+8\left(\tilde{g}+\tfrac{\pi\tilde{a}^2}{2}\right)\tilde{\lambda}^2\right)\sqrt{\delta^2-\tilde{\lambda}^2}~.
\end{equation}
Here, $c_2$, the second moment of $u(\tilde{\lambda})$, is determined by a self-consistency condition. This solution enables us to locate a phase transition, i.e.\ a curve $\CC$ in the $r$-$g$-plane, at which the eigenvalue density $u(\tilde{\lambda})$ becomes negative. This curve is found to be given by 
\begin{equation}
\CC^\pm\ =\ \frac{\pi}{32}\left(-63\tilde{a}^2+16\tilde{r}^2\pm(4\tilde{r}-\tilde{a})\sqrt{16 \tilde{r}^2-8\tilde{a}\tilde{r}-95\tilde{a}^2}-8\tilde{a}\tilde{r}\right)~.
\end{equation}
Various arguments suggest to identify the turning point of this curve with the triple point of the phase diagram predicted numerically. The location of this point for $\tilde{a}=1$, $(r,g)\approx(-2.7,0.25)$, is in good agreement with the numerical result $(r,g)=(-2.3\pm0.2,0.52\pm0.02)$ and justifies further this identification.

As mentioned in the introduction, modifications to the action of fuzzy scalar field theory are necessary, if one wants to employ fuzzy spaces to regularize $\phi^4$-theory on the plane. The modification proposed in \cite{Dolan:2001gn} takes into account a wave function renormalization and -- in its simplest form -- reads as 
\begin{equation}\label{ActionGenMod}
\tilde{S}\ =\ \gamma\tr\left(a \Phi (C_2+\kappa C_2C_2)\Phi+r\,\Phi^2+g\,\Phi^4\right)~.
\end{equation}
This modification is readily treated in our formalism, as it amounts to 
\begin{equation}
 K_{ab} \ \rightarrow\  \check{K}_{ab}\ :=\ K_{ab}+\kappa K_{ac}K_{cb}\eand
a \ \rightarrow\ \tilde{\check{a}}\ =\ a(1+\tfrac{2}{3}\tilde{\kappa})~.
\end{equation}
Increasing $a$ now moves the turning point of $\CC$ -- and thus the triple point -- off to infinity. We thus confirmed that introducing $\kappa$ has the desired effect.

\section{Summary and future directions}

Altogether, we achieved the following: We formulated a generalized character expansion technique for the treatment of fuzzy scalar field theory. This technique yields exact results at any order in a perturbative expansion of the kinetic term and can be evaluated -- in principle -- straightforwardly. It is worth stressing that this approach is directly applicable to field theories on the fuzzy sphere with more general potential and it is readily adapted to scalar field theories on other fuzzy spaces. We used this expansion technique to reformulate scalar field theory on the fuzzy sphere as a multitrace matrix model. The results of the approximation look promising and motivate further studies.

An important task for future work will be to explore the full set of one- and two-cut solutions as well as their r{\^o}le in the explanation of the phase diagram. The qualitative effects of higher order corrections have to be taken into account, if the full analysis at second order should not suffice to explain the phase structure. Obviously, our technique should also be applied to the probably more interesting case of four dimensional theories. Finally, one might wonder whether there is a connection between the multitrace matrix model we found from fuzzy scalar field theory and the original motivation for studying multitrace matrix models \cite{Das:1989fq}: the definition of $c>1$ string theories.



\end{document}